\documentclass[a4paper,11pt]{article}
\pdfoutput=1 % if your are submitting a pdflatex (i.e. if you have
\usepackage{jheppub} % for details on the use of the package, please
\usepackage[T1]{fontenc} % if needed
\usepackage{amssymb}
\usepackage{slashed}
\usepackage{graphicx}
\usepackage{epstopdf}
\usepackage{diagbox}
\title{\boldmath A novel strategy for searching for $C\!P$ violations in the baryon sector}

\author[a]{Zhen-Hua Zhang}
\author[b]{Xin-Heng Guo}

\affiliation[a]{School of Nuclear Science and Technology, University of South China, Hengyang, Hunan, 421001, China.}
\affiliation[b]{College of Nuclear Science and Technology, Beijing Normal University, Beijing, 100875, China.}

\emailAdd{zhangzh@usc.edu.cn}
\emailAdd{xhguo@bnu.edu.cn}

\abstract{
Despite the large baryon-anti-baryon asymmetry in the observable Universe, the closely related phenomenon -- the violation of the combined charge and parity symmetry ($C\!P$V) -- has not been observed in the baryon sector in laboratories.
In this paper, a new strategy for searching for $C\!P$V in heavy hadron multi-body decays is proposed, in which a set of novel observables measuring $C\!P$V in such decays -- the partial wave $C\!P$ asymmetries (PW$C\!P$As) -- are introduced.
This strategy is model-independent and applicable to multi-body decays of heavy hadrons with arbitrary spin configurations in both initial and final states, and with any number of particles in the final state.
It is especially applicable for $C\!P$V investigations in multi-body decays of heavy baryons.
As applications of this strategy, we suggest to measure the PW$C\!P$As in some decay channels of bottom baryons such as $\Lambda_b^0\to p \pi^-\pi^+\pi^-$, $\Lambda_b\to p K^- \pi^+\pi^-$, $\Lambda^0_b\to p\pi^-K^+K^-$, $\Lambda_b^0\to\Lambda K^+\pi^-$, and $\Lambda_b^0\to p  \pi^- K_s$.}

\begin{document}
\maketitle
\flushbottom

%%%%%%%%%%%%%%%%%%%%%%%%%
\section{\label{sec:introduction}Introduction}
%%%%%%%%%%%%%%%%%%%%%%%%%
The violation of the symmetry of the combined charge-parity ($C\!P$) transformation, as a phenomenon closely related to the matter-antimatter asymmetry in the Universe \cite{Sakharov:1967dj}, was first discovered in the neutral kaon system \cite{Christenson:1964fg}, and
is accommodated in the Standard Model (SM) of particle physics by the
Cabibbo-Kobayashi-Maskawa (CKM) mechanism that describes the transitions between up- and down-type quarks\cite{Cabibbo:1963yz,Kobayashi:1973fv}.
$C\!P$ violation ($C\!P$V) in hadron decays is described by the asymmetry between the particle and antiparticle decay rates, which has been observed in many decay channels of heavy mesons \cite{Aubert:2001nu,Abe:2001xe,Aaij:2013iua,Aaij:2013sfa,Aaij:2019kcg}.
On the other hand, despite the large baryon-anti-baryon asymmetry in the observable Universe,
no $C\!P$V has ever been observed in the baryon sector in laboratories \cite{Aaij:2016cla,Aaij:2017xva,Aaij:2018tlk,Aaij:2018lsx,Aaij:2019mmy,Aaij:2020wil}.

One reason for the negative result in searching for $C\!P$Vs in the baryon sector is the substantially lower statistics compared with the meson case.
One of the latest examples for the $C\!P$V measurements in bottom baryon decays can be found in Ref. \cite{Aaij:2021oow}, from which one can clearly see that the statistics for the bottom baryon case is indeed much lower than that in the bottom meson case (see, for example, Ref. \cite{LHCb:2020byh}).
So although $C\!P$ asymmetries ($C\!P$As) of a few or several tens of percent is expected in some decay channels of heavy baryons \cite{Hsiao:2014mua}, no significant evidence of $C\!P$V has yet been found experimentally in the baryon sector.
%Therefore, in order to discover $C\!P$V with the fewer data in baryon decays, a more efficient strategy for $C\!P$V hunting in the baryon sector becomes urgent.

The rich resonance structures in multi-body decays of heavy hadrons provide us opportunities for $C\!P$V searching in these decay channels, as the interference between different intermediate resonances may result in large regional $C\!P$As (r$C\!P$As) distributed in the phase space, although the overall $C\!P$A may be small due to cancellations among different parts of the phase space.
In fact, it has been observed that for some three-body decay channels of $B$ mesons such as $B^+\to \pi^+\pi^-\pi^+$ and $B^+\to K^+\pi^-\pi^+$, very large r$C\!P$As are localized in certain small parts of the phase space, thanks to the relatively larger statistics \cite{Aaij:2013sfa}.
Some of these r$C\!P$As can indeed be explained by the interference between a resonance and another one nearby via the introduction of a non-perturbative relative strong phase between their corresponding amplitudes \cite{Bediaga:2009tr,Dedonder:2010fg,Zhang:2013oqa}.
Large r$C\!P$As associated with the interference between close resonances are also expected in multi-body decays of heavy baryons.
However, the aforementioned low statistics problem leads to the direct measurements of the r$C\!P$As for multi-body decays of heavy baryons suffering from large uncertainties and hence no $C\!P$V has been established in heavy baryon decays.
Even though one can partly fix the low statistic issue by merging small bins in the phase space region into larger ones, the r$C\!P$As of the enlarged bins are usually much smaller since there are almost always cancellations among
these aforementioned small bins when obtaining the r$C\!P$As of the enlarged bins.
Therefore, besides the direct measurements of the r$C\!P$As or the overall $C\!P$As, a more efficient and systematic method for $C\!P$V hunting is thus urgent for multi-body decays of heavy baryons.

Motivated by the above discussion, in this paper, we will introduce a set of new $C\!P$V observables -- the Partial Wave $C\!P$ Asymmetries -- associated with the intermediate resonances in the multi-body decays of heavy hadrons.
It will be shown that the Partial Wave $C\!P$ Asymmetries can make use of the data in a more efficient way, and thus, can potentially overcome the low statistics problem associated with the heavy baryon decays.
It is possible that with this new strategy, the establishment of $C\!P$V in the baryon sector may be hastened.

This paper is organized as follows.
In Sec. \ref{sec:PWCPA}, we introduce the set of observables, the Partial Wave $C\!P$ Asymmetries, and discuss its measurement method and its advantages comparing with other observables or techniques for $C\!P$V measurements.
In Sec. \ref{sec:SR}, we focus mainly on the situation of the interference of nearby resonances, in which the non-perturbative effect of strong interaction could amplify the $C\!P$V signal.
We briefly discuss the correlation of the underline dynamics for $C\!P$V with this newly introduced observables for this situation.
An set of important Selection Rules which tell weather the interference effect are present or absent in certain Partial Wave $C\!P$ Asymmetries are also presented.
In Sec. \ref{sec:application}, we apply the newly introduced observables to the four-body decay process, $\Lambda_b^0\to p \pi^-\pi^+\pi^-$.
In the last section, we briefly give the conclusion.

%%%%%%%%%%%%%%%%%%%%%%%%%
\section{\label{sec:PWCPA}\boldmath Partial Wave $C\!P$ Asymmetries}
%%%%%%%%%%%%%%%%%%%%%%%%%
Without lose of generality, consider a multi-body decay, $H\to h_1 h_2 h_3\cdots h_n$, where $H$ is a heavy hadron, while $h_k$ ($k=1,2,3,\cdots,n$) are light ones.
We will focus on the region of the phase space where an intermediate resonance $R_a$ dominates, i.e., the invariant mass squared of the $h_1h_2$ system, $s_{12}$, lies between $(m_{R_a}-\alpha_{R_a})^2$ and $(m_{R_a}+\beta_{R_a})^2$, with $m_{R_a}$ the mass of $R_a$, and $\alpha_{R_a}$ and $\beta_{R_a}$ of the same other as the decay width of $R_a$.
Then, this multi-body decay will be dominated by the cascade decay $H\to R_a h_3\cdots h_n$, $R_a\to h_1h_2$.
Assuming that the initial particle $H$ is produced unpolarized, the differential decay width can then be expressed as
\begin{equation}%\label{}
 d\Gamma\propto
\overline{\left|\mathcal{M}\right|^2}dc_{\theta_1^\ast},
\end{equation}
after summing over the helicities (or the z-components of the spins) of both the initial and the final particles, and integrating over the phase space variables ---$s_{12}$ is integrated from $(m_{R_a}-\alpha_{R_a})^2$ to $(m_{R_a}+\beta_{R_a})^2$ for the phase space integration--- except $c_{\theta_1^\ast}$ ,
where $\theta_{1}^\ast$ is the relative angle between the momentum of $h_1$ and that of $H$ in the rest frame of the $h_1h_2$ system, and $c_{\theta_{1}^\ast}$ is its cosine, $\overline{\left|\mathcal{M}\right|^2}$ is the spin-averaged and phase space-integrated (except $c_{\theta_1^\ast}$) square of the decay amplitude, which can be further expanded with the aid of the Legendre's polynomials as
\begin{equation}
\label{eq:AmpSquare}
  \overline{\left|\mathcal{M}\right|^2}\propto\sum_{j=0}^{\infty}w^{(j)}P_{j}(c_{\theta_{1}^\ast}),
\end{equation}
where the coefficients $w^{(j)}$ can be expressed as
\begin{equation}
w^{(j)}\propto \int_{-1}^{1}\overline{\left|\mathcal{M}\right|^2} P_{j} dc_{\theta_{1}^*},
\end{equation}
according to the orthogonality properties of $P_j$s.
Since this decay process is a weak one, $C\!P$V may show up as the difference between $w^{(j)}$ and $\bar{w}^{(j)}$, where $\bar{w}^{(j)}$ corresponds to ${w}^{(j)}$ for the $C\!P$ conjugate process $\bar{H}\to \bar{h}_1 \bar{h}_2 \bar{h}_3 \cdots\bar{h}_n$.
One can then introduce a set of observables, which will be called as the Partial Wave $C\!P$ Asymmetries (PW$C\!P$As), and for the $j$th-wave:
\begin{equation}%\label{}
  A_{CP}^{(j)}\equiv \frac{w^{(j)}-\bar{w}^{(j)}}{ w^{(j)}+\bar{w}^{(j)}}.
\end{equation}

The PW$C\!P$As are experiment-friendly observables, which can be extracted easily by fitting the data with Eq. (\ref{eq:AmpSquare}).
Another equivalent but more straightforward method is described as follows.
According to the orthogonality property of $P_j$ one can see that different events collected in the data do not contribute  to $w^{(j)}$ equally.
Instead, they contribute with a weight which is proportional to $P_{j}(c_{\theta_{1}^\ast})$.
For an event labeled as $k$, one can easily determine its corresponding $c_{\theta_{1}^\ast}$, which will be denoted as $c_{\theta_{1,k}^\ast}$.
Accordingly, one can introduce the so called $P_j$-weighted event yield,
\begin{equation}%\label{}
  \mathcal{N}_{j-\text{weighted}}\equiv \sum_{k} P_{j} (c_{\theta_{1,k}^\ast}),
\end{equation}
which is proportional to $w^{(j)}$.\footnote{
Equivalently, to obtain $ \mathcal{N}_{j-\text{weighted}}$, one can also firstly divide the range of $c_{\theta_{1}^\ast}$, from $-1$ to $1$,  into $M$ small intervals, with each interval corresponding to an average value of $c_{\theta_{1}^\ast}$, which will be denoted as $c_{\theta_{1,m}^\ast}$, $(m=1,2,\cdots,M)$.
Secondly, count the event number in each interval, which will be denoted as ${N}_m$ for the interval $m$.
Then, one can easily see that the $P_j$-weighted event yields $\mathcal{N}_{j-\text{weighted}}$ can be expressed as
$%\begin{equation}%\label{}
 \mathcal{N}_{j-\text{weighted}}=\sum_m {N}_m P_j(c_{\theta_{1,m}^\ast}). %\nonumber
$ %\end{equation}
This method will be used to obtain the PW$C\!P$A in the simulation in Sec. \ref{sec:application}.
}
The $P_j$-weighted event yield for the $C\!P$ conjugate process $\bar{\mathcal{N}}_{j-\text{weighted}}$ can be obtained in the same way.
With the above $P_j$-weighted event yields $\mathcal{N}_{j-\text{weighted}}$ and $\bar{\mathcal{N}}_{j-\text{weighted}}$, one can easily obtain the experimental values of the PW$C\!P$As through
\begin{equation}\label{eq:ACPjexp}
  A_{CP}^{(j),{\text{exp}}}= \frac{\mathcal{N}_{j-\text{weighted}}-\bar{\mathcal{N}}_{j-\text{weighted}}}{\mathcal{N}_{j-\text{weighted}}+\bar{\mathcal{N}}_{j-\text{weighted}}}.
\end{equation}

From the measurement methods proposed above, one can see that with the newly introduced observables PW$C\!P$As one can make use of the data in a more efficient way, comparing with the r$C\!P$As which are widely measured in multi-body decays of $B$ mesons currentle.
This is because instead of dividing the value range of $c_{\theta_{1}^\ast}$ into small bins and making use of only the events in each bins when measuring the r$C\!P$As, each of the PW$C\!P$As makes use of all the events distributed in the {\it whole} range of $c_{\theta_{1}^\ast}$ from $-1$ to 1.
This can be seen clearly when the integration over $c_{\theta_{1}^*}$ from $-1$ to 1 is performed based on the orthogonality properties of $P_j$, indicating the usage of the events over the whole range of $c_{\theta_{1}^*}$.
This can also be seen in more detail from the measurement method proposed above.
Making use of all the events distributed in the whole range of $c_{\theta_{1}^\ast}$ is very important when statistics is not large enough, as in the case of heavy baryon decays.
Another important feature of the PW$C\!P$As which can be seen from the above discussion is that, they provide a more efficient model-independent approach for hunting for $C\!P$V than the currently used amplitude analysis technique \cite{Dalitz:1953cp} in multi-body decays of heavy hadrons, in which model dependence could potentially arise \cite{Aaij:2019jaq}.

%%%%%%%%%%%%%%%%%%%%%%%%%
\section{\label{sec:SR}\boldmath Contribution of the interference of nearby resonances to PW$C\!P$A: underlining dynamics and Selection Rules}
%%%%%%%%%%%%%%%%%%%%%%%%%
The PW$C\!P$As $A_{CP}^{(j)}$ are in fact $C\!P$V observables for the decay $H\to R_a h_3\cdots h_n$, if $R_a$ is the {\it only} dominate resonance in the phase space region, $(m_{R_a}-\alpha_{R_a})^2<s_{12}<(m_{R_a}+\beta_{R_a})^2$.
For example, $A_{CP}^{(0)}$ is nothing but the $C\!P$A parameter for the decay $H\to R_a h_3\cdots h_n$.\footnote{
Of course, $A_{CP}^{(0)}$ is also the r$C\!P$A for the whole $R_a$-dominated phase space region,  $(m_{R_a}-\alpha_{R_a})^2<s_{12}<(m_{R_a}+\beta_{R_a})^2$.}
However, $C\!P$V is usually relatively small in the single-resonance-dominance situation.
Hence, interference with a second resonance is required to generate larger $C\!P$V.
In this situation, the decay amplitude can then be approximated by the sum of the two cascade decays, $H\to R_a h_3\cdots h_n$, $R_a\to h_1 h_2$, and $H\to R_b h_3\cdots h_n$, $R_b\to h_1 h_2$, in the phase space region $(m_{R_a}-\alpha_{R_a})^2<s_{12}<(m_{R_a}+\beta_{R_a})^2$.
In this situation, the PW$C\!P$As are no longer just $C\!P$V observables for $H\to R_a h_3\cdots h_n$.
Instead, they will also contain contribution from the resonance $R_b$, and moreover, that from the interference between the two resonances $R_a$ and $R_b$.
The interference of nearby resonances has great impact on $C\!P$V in multi-body decays of bottom hadrons, either through r$C\!P$As, as has been observed in some three-body decays of $B$ meson, and/or through PW$C\!P$As, which will be seen in more details in what follows.

It is crucially important to find out the correlation between the origin of $C\!P$V in the PW$C\!P$As, $A_{CP}^{(j)}$, and the underlining dynamics.
To achieve this, one needs to substitute the decay amplitudes into $\overline{\left|\mathcal{M}\right|^2}$.
After some algebra, one arrive at the
expression of $w^{(j)}$ as
\begin{equation}\label{eq:wj}
w^{(j)}=\left\langle\frac{\mathcal{S}_{aa}^j \mathcal{W}_{aa}^j}{|s_{R_a}|^2}\right\rangle+\left\langle\frac{\mathcal{S}_{bb}^j \mathcal{W}_{bb}^j}{|s_{R_b}|^2}\right\rangle+\left\langle2 \Re\left(\frac{\mathcal{S}_{ab}^{j}\mathcal{W}_{ab}^j}{s_as_b^\ast}\right)\right\rangle,
\end{equation}
where $s_{R_a/R_b}$ are the denominators of the Breit-Wigner propagators and take the form $s_{R_a/R_b}=s_{12}-m_{R_a/R_b}^2+im_{R_a/R_b}\Gamma_{R_a/R_b}$, the notation ``$\langle\cdots\rangle$'' represents the integral over the phase space except $c_{\theta_1^\ast}$, and
\begin{equation}
\label{eq:wXY}
\mathcal{W}_{XY}^j=\sum_{\sigma}(-)^{\sigma-j} \langle j_{X}j_{Y}-\sigma\sigma|j0\rangle \sum_{m_z\lambda_3\cdots}\mathcal{M}^{Jm_z,X}_{\sigma\lambda_3\cdots}{\mathcal{M}}^{Jm_z,Y\!*}_{{\sigma}{\lambda}_3\cdots},
\end{equation}
\begin{equation}
%\label{Eq:wXorY}
\mathcal{S}_{XY}^{j}=\sum_{{\lambda}_1'{\lambda}_2'}\left.(-)^{j-\lambda'}\langle j_{X}j_{Y}-\lambda'\lambda'|j0\rangle\mathcal{G}^{j_X}_{\lambda_1'\lambda_2'}\mathcal{G}^{j_Y*}_{{\lambda}_1'{\lambda}_2'}\right|_{\lambda'=\lambda_1'-\lambda_2'},
\end{equation}
where $X,Y=a,b$, $j_{a/b}$ are the spins of $R_{a/b}$, $\mathcal{M}^{Jm_z,a/b}_{\sigma\lambda_3\cdots}$ are the covariant decay amplitudes for the weak decays $H \to R_{a/b} h_3\cdots h_n$ with $\sigma, \lambda_3,\cdots$ being the helicities  of $R_{a/b}, h_3,\cdots$ defined in the rest frame of $H$ , respectively, $\mathcal{G}^{j_{a/b}}_{\lambda_1'\lambda_2'}$ are the helicity decay amplitudes \cite{Jacob:1959at} for the strong decay processes $R_{a/b}\to h_1 h_2$ in the rest frame of the $h_1h_2$ system with $\lambda_{1(2)}'$ being the helicities of $h_{1(2)}$ in the same frame, and the notation $\langle\cdots|\cdots\rangle$'s are the Clebsch-Gordan coefficients.
Among the three terms in Eq. (\ref{eq:wj}), the first two represent the contributions from the resonances $R_a$ and $R_b$ alone, respectively, while the last one represents the effect of the interference between $R_a$ and $R_b$.

As of aforementioned, the PW$C\!P$As get their contributions from the difference between $w^{j}$ and $\bar{w}^{j}$.
This difference is originated from the interference of amplitudes with different weak and strong phases.
For example, for the case of bottom hadron decay, each of the weak decay amplitudes $\mathcal{M}^{Jm_z,a/b}_{\sigma\lambda_3\cdots}$ can be further divided into a tree and penguin amplitudes, $\mathcal{M}^{Jm_z,a/b}_{\sigma\lambda_3\cdots}=\mathcal{M}^{Jm_z,a/b,\text{tree}}_{\sigma\lambda_3\cdots}+\mathcal{M}^{Jm_z,a/b,\text{penguin}}_{\sigma\lambda_3\cdots}$.
All the three terms in Eq. (\ref{eq:wj}) could contribute to the PW$C\!P$As.
The contributions to the PW$C\!P$As of the first two terms in Eq. (\ref{eq:wj}) come from the weak decay process $H\to R_{a/b} h_3\cdots h_n$ {\it alone}, i.e., the interference of the tree and penguin amplitudes for the decay via the same resonances $R_a$ or $R_b$.
These two terms are proportional to the sine of the strong phase difference between the amplitudes of the tree and penguin operators for {\it the same} resonances $R_a$ or $R_b$, respectively, which is usually small unless some particular mechanism enters to generate a large phase difference.\footnote{
Note that if there were only one resonance $R_a$  which dominates, $w^{j}$ would only contain the first term.
In this situation, the PW$C\!P$As are reduced to
\begin{equation}%\label{}
  A_{CP}^{(j)}=\frac{\mathcal{W}_{aa}^j-\bar{\mathcal{W}}_{aa}^j}{ \mathcal{W}_{aa}^j+\bar{\mathcal{W}}_{aa}^j}, \nonumber\\
\end{equation}
from which one can see that the PW$C\!P$As are indeed observables for the decay $H\to R_{a} h_3\cdots h_n$ and they contain only the weak decay amplitudes $\mathcal{M}^{Jm_z,a}_{\sigma\lambda_3\cdots}$.
}
On the other hand, the last term in Eq. (\ref{eq:wj}) represents the interference between the two resonances $R_a$ and $R_b$.
Its contribution to the PW$C\!P$As is proportional to the sine of the strong phase difference between the amplitude through the resonance $R_a$ and that through $R_b$, i.e., tree and penguin amplitudes from different resonances.
The strong phase difference between different resonances could be large because of the non-perturbative strong interaction effects, permitting the existence of large r$C\!P$As and/or PW$C\!P$As.
In fact, large r$C\!P$As which are correlated with the aforementioned interference of different resonances have been observed in some three-body decays of $B$ meson.
Similarly, the presence of the last term in $w^{j}$ in Eq. (\ref{eq:wj}) indicates that it can contribute to the PW$C\!P$As, resulting in the PW$C\!P$As which are large enough to be potentially detectable.

With the aid of the properties of the Clebsch-Gordan coefficients and the parity-conservation requirement for the processes $R_{a/b}\to h_1 h_2$, it is easy to show that the interference and non-interference terms satisfy the following Selection Rules (SRs):
the interference terms show up only when 1) $j=|j_a-j_b|,\cdots, j_a+j_b$ and 2) $(-)^{j}\pi_a\pi_b$ is positive, where $\pi_a$ and $\pi_b$ are the parities of $R_a$ and $R_b$, respectively;
while the non-interference terms show up only when 1) $j$ is even, and 2) $j=0,\cdots, 2j_{a/b}$.
In Table \ref{tab:IntHalfIntSR}, we list the values of $j$ for the interference and non-interference terms appear in $w^{(j)}$ and $\bar{w}^{(j)}$ according to the SRs for some spin-parity configurations of the resonances $R_a$ and $R_b$.

%%%%%%%%%%%%%%%%%%%%%%%%%
\section{\label{sec:application}Applications to baryon decays}
%%%%%%%%%%%%%%%%%%%%%%%%%
The previously introduced PW$C\!P$As provide a systematic way to investigate the underlining dynamics of $C\!P$V in multi-body decays of heavy hadrons.
We propose to search for $C\!P$V first through the measurements of the PW$C\!P$As in multi-body decay channels via the transitions $b\to u\bar{u}d$ or $b\to u \bar{u}s$, for which large weak phases are expected.
Candidates include $\Lambda_b^0\to p \pi^-\pi^+\pi^-$, $\Lambda_b\to p K^- \pi^+\pi^-$, $\Lambda^0_b\to p\pi^-K^+K^-$, $\Lambda_b^0\to\Lambda K^+\pi^-$, and $\Lambda_b^0\to p K_s \pi^-$, et.al..\footnote{
In fact, $C\!P$V has been investigated in decay channels such as $\Lambda_b^0\to p \pi^-\pi^+\pi^-$ and $\Lambda_b\to p K^- \pi^+\pi^-$, where $C\!P$As associated with the triple product asymmetry \cite{Gronau:2015gha,Durieux:2015zwa} were measured, and no $C\!P$V was established \cite{Aaij:2016cla,Aaij:2018lsx,Aaij:2019mmy}.}

Take the decay channel $\Lambda_b^0\to p \pi^-\pi^+\pi^-$ as an example.
The rich resonance structure of this decay channel makes it also a perfect channel for the measurement of PW$C\!P$As.
The dominance of the resonances such as $\Delta^{++}(1232)$ and $\rho^0(770)$ has been observed by LHCb in this decay channel \cite{Aaij:2018lsx,Aaij:2019mmy}.
Based on a simple isospin symmetry analysis one can deduce that the resonance $\Delta^{0}(1232)$ should also dominate.
Meanwhile, since the nearby resonance $N^0(1440)$ has a width as large as $350$ MeV, the interference effect between the two resonances, $\Delta^0(1232)$ and $N^0(1440)$ (referred as $\Delta^0$ and $N^0$ respectively hereinafter), around the vicinity of $\Delta^0$, could be large.
As a consequence, the interference between the two cascade decays, $\Lambda_b^0\to N^0(\to p \pi^-)\pi^+\pi^-$ and $\Lambda_b^0\to \Delta^0(\to p \pi^-)\pi^+\pi^-$, could potentially generate PW$C\!P$As which are large enough to be detected.

Since the spins-parities of these two resonances are ${\frac{1}{2}}^+$ and ${\frac{3}{2}}^+$, respectively, from the SRs (and also, Table \ref{tab:IntHalfIntSR}) one can deduce that the non-interference terms show up for $j=0, 2$, while the interference term only shows up for $j=2$.
This means that the potentially large $C\!P$V induced by the interference of these two nearby resonances $\Delta^0$ and $N^0$ are hence only embedded in the PW$C\!P$A $A_{C\!P}^{(2)}$, with no such contribution to the r$C\!P$A $A_{CP}^{(0)}$ at all.
The measurement of the r$C\!P$A $A_{C\!P}^{(0)}$, as is conventionally done, would miss the interference-induced potentially large $C\!P$V.
It could only be possible to find $C\!P$V corresponding to the interference effect between $N^0$ and $\Delta^0$ through the measurement of $A_{C\!P}^{(2)}$.

To simplify the numerical estimation, while confining the invariant mass square of the $p\pi^-$ system, $s_{p\pi^-}$, to be around $m_\Delta^2$,
 we further constrain the invariant mass of the remaining $\pi^+\pi^-$ pair to be around the $\rho^0(770)$ (referred as $\rho^0$ hereinafter), which will not reduce the statistics significantly due to the dominance of $\rho^0$.
In this situation, the decay process is dominated by two coherent cascade decays $\Lambda_b^0\to N^0(\to p \pi^-)\rho^0(\to\pi^+\pi^-)$ and $\Lambda_b^0\to \Delta^0(\to p \pi^-)\rho^0(\to\pi^+\pi^-)$.
Via the generalized factorization approach, the amplitudes of $\Lambda_b^0\to \Delta^0\rho^0$ and $\Lambda_b^0\to N^0\rho^0$ can both be parameterized as:
\begin{equation}%\label{}
  \mathcal{A}_{\Lambda_b^0\to X\rho^0}\propto \alpha_X\langle X |\overline{u}\slashed{\varepsilon}^\ast(1-\gamma_5)b|\Lambda_b^0\rangle,
\end{equation}
where $X=\Delta^0$ or $N^0$, $\langle X |\overline{u}\slashed{\varepsilon}^\ast(1-\gamma_5)b|\Lambda_b^0\rangle$ is the matrix element for the transition $\Lambda_b^0\to X$ with $\varepsilon$ the polarization vector of $\rho^0$, and
\begin{equation}%\label{}
  \alpha_X=V_{ub}V_{ud}^\ast a_{2,X}-V_{tb}V_{td}^\ast a_{4,X},
\end{equation}
with $V_{ub}$, $V_{ud}$, $V_{tb}$, and $V_{td}$ the CKM matrix elements, $a_{i,X}=c_{i}^{\text{eff}}+c_{i-1}^{\text{eff}}/N_C^{\text{eff},X}$ for $i=2,4$ ($c_{i}^{\text{eff}}$ and $N_C^{\text{eff},X}$ are the effective Wilson coefficients and color number, respectively).
The amplitudes for $\Lambda_b^0\to N^0(\to p \pi^-)\rho^0(\to\pi^+\pi^-)$ and $\Lambda_b^0\to \Delta^0(\to p \pi^-)\rho^0(\to\pi^+\pi^-)$ can be respectively expressed in the following helicity forms:
\begin{equation}%\label{}
  \mathcal{F}_{\sigma}^\Delta\mathcal{G}_{\lambda_p'}^\Delta\sim \alpha_{\Delta}[c_{P-\text{conserve}}^\Delta(2\sigma)+c_{P-\text{violate}}^\Delta]{(2\lambda_p)}e^{i\delta_\Delta},
\end{equation}
\begin{equation}%\label{}
  \mathcal{F}_{\sigma}^N\mathcal{G}_{\lambda_p'}^N\sim \alpha_{N}[c_{P-\text{conserve}}^N+c_{P-\text{violate}}^N(2\sigma)]e^{i\delta_N},
\end{equation}
where we have also written the weak decay amplitudes in the helicity form, $\mathcal{F}_{\sigma}^{\Delta/N}$, the parameters $c_{P-\text{conserve/violate}}^{\Delta,N}$ contain the form factors for the transitions $\Lambda_b^0\to \Delta^0/N^0$, the decay amplitudes for the strong decay processes $\Delta/N\to p\pi^-$, and some other common factors corresponding to the $\rho^0$ resonance which also depend on the phase space variables.

\begin{figure}
  \centering
  \includegraphics[width=\textwidth]{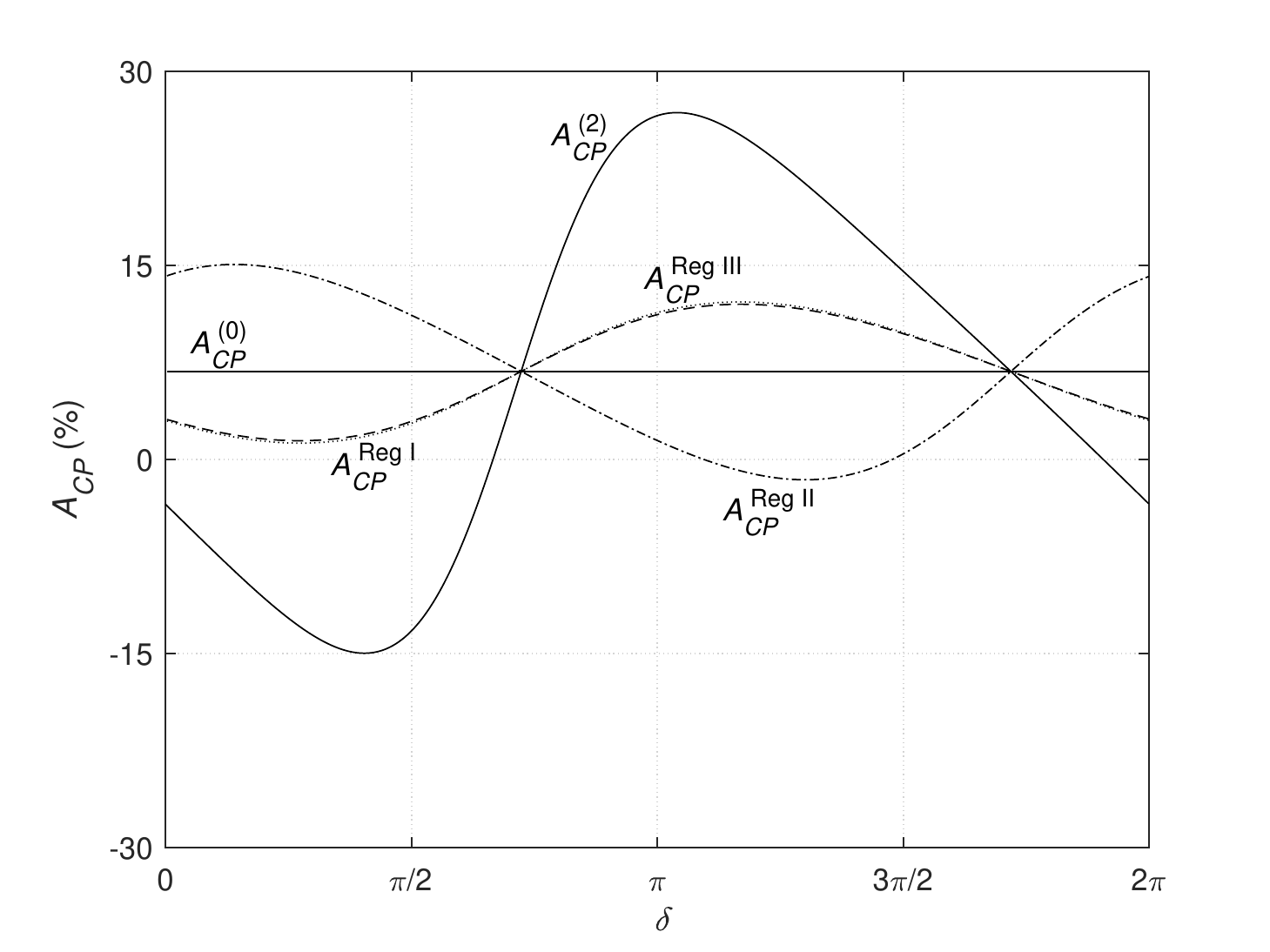}\\
  \caption{The PW$C\!P$A $A_{C\!P}^{(2)}$ (solid curvy line) for $\Lambda_b^0\to p\pi^-\pi^+\pi^-$ near the resonance $\Delta^0(1232)$ as a function of the strong phase $\delta$. The regional $C\!P$ asymmetry $A_{C\!P}^{(0)}$ (solid straight line), $A_{C\!P}^{\text{Reg~I}}$ (dotted line), $A_{C\!P}^{\text{Reg~II}}$ (dash-dotted line), and $A_{C\!P}^{\text{Reg~III}}$ (dashed line)  are also shown for comparison. The difference between $A_{C\!P}^{\text{Reg~I}}$ and $A_{C\!P}^{\text{Reg~III}}$ is very tiny. Other PW$C\!P$As $A_{C\!P}^{(1)}$ and $A_{C\!P}^{(3)}$ are not shown due to the reason explained in the text. The invariant mass squared $s_{p\pi}$ is integrated from $(m_\Delta-\Gamma_\Delta)^2$ to $(m_\Delta+\Gamma_\Delta)^2$.}\label{fig:PWCPA}
\end{figure}

Although the strong coupling constants can be extracted from the corresponding branching ratios, the weak transition form factors are not available.
Moreover, the relative strong phase $\delta\equiv \delta_N-\delta_\Delta$ is not available either due to its non-perturbative nature.
These prevent us from an accurate prediction for the PW$C\!P$As in $\Lambda_b^0\to p \pi^-\pi^+\pi^-$.
Nonetheless, for the purpose of illustrating the behaviours of the PW$C\!P$As, we simply set all the $\langle c_{P-\text{conserve/violate}}^{\Delta,N}\rangle$ to be the same.
Then, the PW$C\!P$As are simply functions of the strong phase $\delta$, which are shown in Fig. \ref{fig:PWCPA}.\footnote{
In drawing this figure, we have used the effective Wilson's coefficients calculated according to Ref. \cite{Deshpande:1994pw} with $q^2/m_b^2=0.3$. Besides, the effective color number is set to be $N_{C}^{\text{eff},\Delta}=2$ and $N_{C}^{\text{eff},N}=2.5$, and the CKM matrix elements are taken from Ref. \cite{Zyla:2020zbs}.}

The dependence of $A_{C\!P}^{(2)}$ on the strong phase $\delta$ can be clearly seen from Fig. \ref{fig:PWCPA}, which is an indication of the presence of the interference term in $A_{C\!P}^{(2)}$, as expected.
On the other hand, the r$C\!P$A, $A_{C\!P}^{(0)}$, which is also presented in this figure for comparison, is independent of $\delta$, indicating the absence of  interference term in $A_{C\!P}^{(0)}$, in line with the SRs constraint.
Moreover, it can be seen that $A_{C\!P}^{(2)}$ is much larger than $A_{C\!P}^{(0)}$ in most regions of $\delta$, indicating a potentially large $C\!P$ asymmetry induced by the interference between $\Delta^0$ and $N^0$.

\begin{figure}
  \centering
  \includegraphics[width=\textwidth]{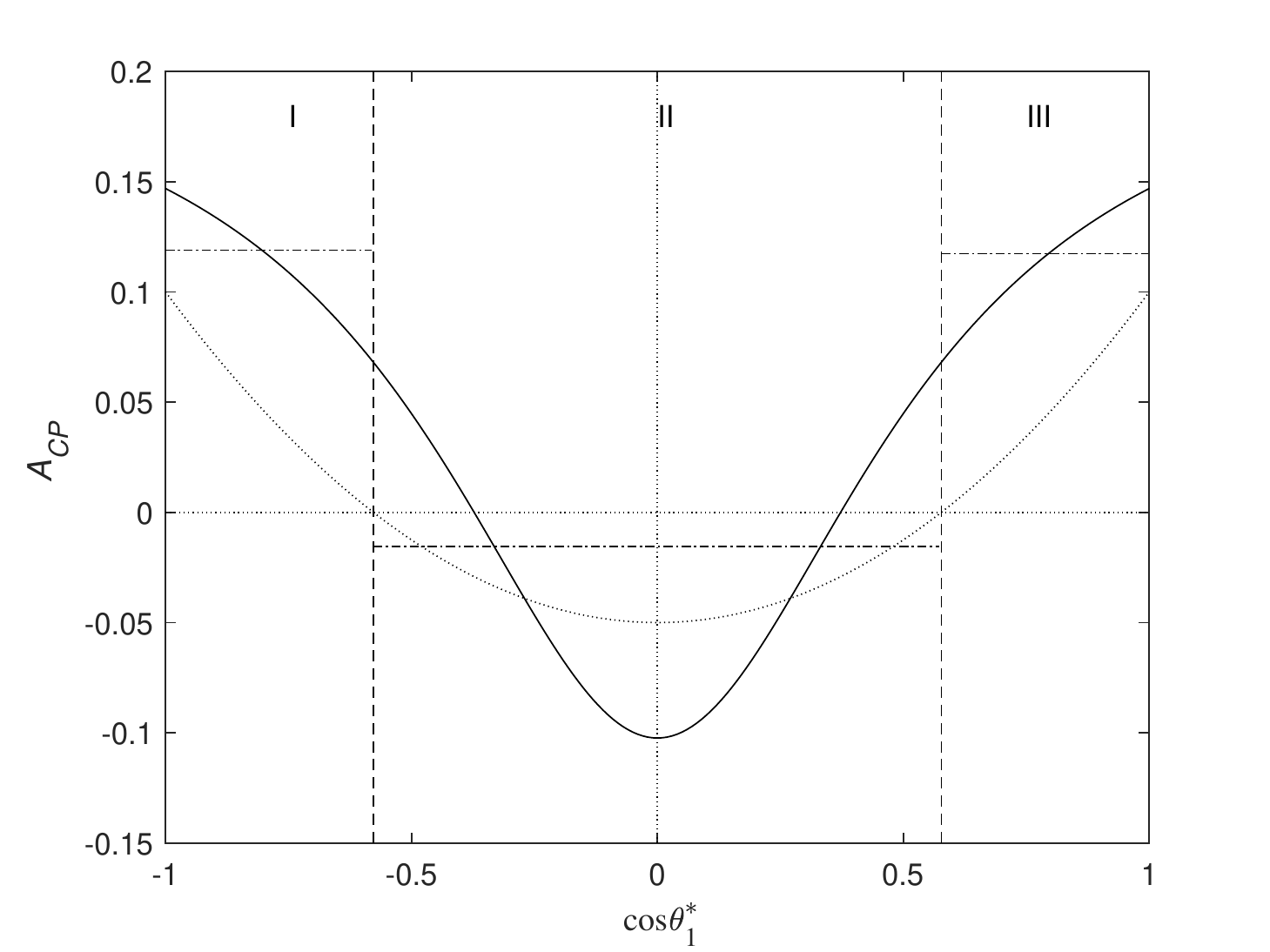}\\
  \caption{Differential $C\!P$A for $\delta=4$ (solid line), r$C\!P$As of the region I, II, and III (dash-dotted line in each region), are shown as a function of $c_{\theta_{1}^\ast}$. The Legendre polynomial $P_2$ (times 0.1) is also shown with dotted line.}\label{fig:RegionalCPA}
\end{figure}

For comparison, the dependences of the differential $C\!P$As on $c_{\theta_{1}^\ast}$ ($\theta_{1}^\ast$ is the angle between the proton and $\rho^0$ for now) for $\delta=4$ are shown in Fig. \ref{fig:RegionalCPA}.
The whole range of $c_{\theta_{1}^\ast}$ is divided into three parts according to the sign of the Legendre polynomial $P_2 (c_{\theta_{1}^\ast})$, which are denoted as I, II, and III, respectively, in Fig. \ref{fig:RegionalCPA}, corresponding to $-1<c_{\theta_{1}^\ast}<-1/\sqrt{3}$, $-1/\sqrt{3}<c_{\theta_{1}^\ast}<1/\sqrt{3}$, and $1/\sqrt{3}<c_{\theta_{1}^\ast}<1$, respectively.
The r$C\!P$As of these three parts $A_{CP}^{\text{Reg~I,~II,~III}}$ are also shown in this figure, from which one can see that the r$C\!P$As tend to change signs in Region I and III comparing with the negative sign in Region II.
This is a bad news for $A_{CP}^{(0)}$, because there will be cancellation between Region II and I + III.
On the contrary, this is a good news for $A_{CP}^{(2)}$, as the extra $P_2$ in $A_{CP}^{(2)}$ results in constructive contributions from all the three parts.
Indeed, the numerical values for the PW$C\!P$As $A_{CP}^{(0)}$ and $A_{CP}^{(2)}$ are calculated to be $A_{CP}^{(0)}=6.8\%$ and $A_{CP}^{(2)}=21.8\%$ for $\delta=4$, respectively.
In order to see more clearly, the  $\delta$-dependence of the r$C\!P$As of Region I, II, and III are also presented in Fig. \ref{fig:PWCPA}, from which is can be seen that for quite a large range of the strong phase $\delta$, the PW$C\!P$A $A_{C\!P}^{(2)}$ is much lager than the r$C\!P$As.

\begin{figure}
  \centering
  \includegraphics[width=\textwidth]{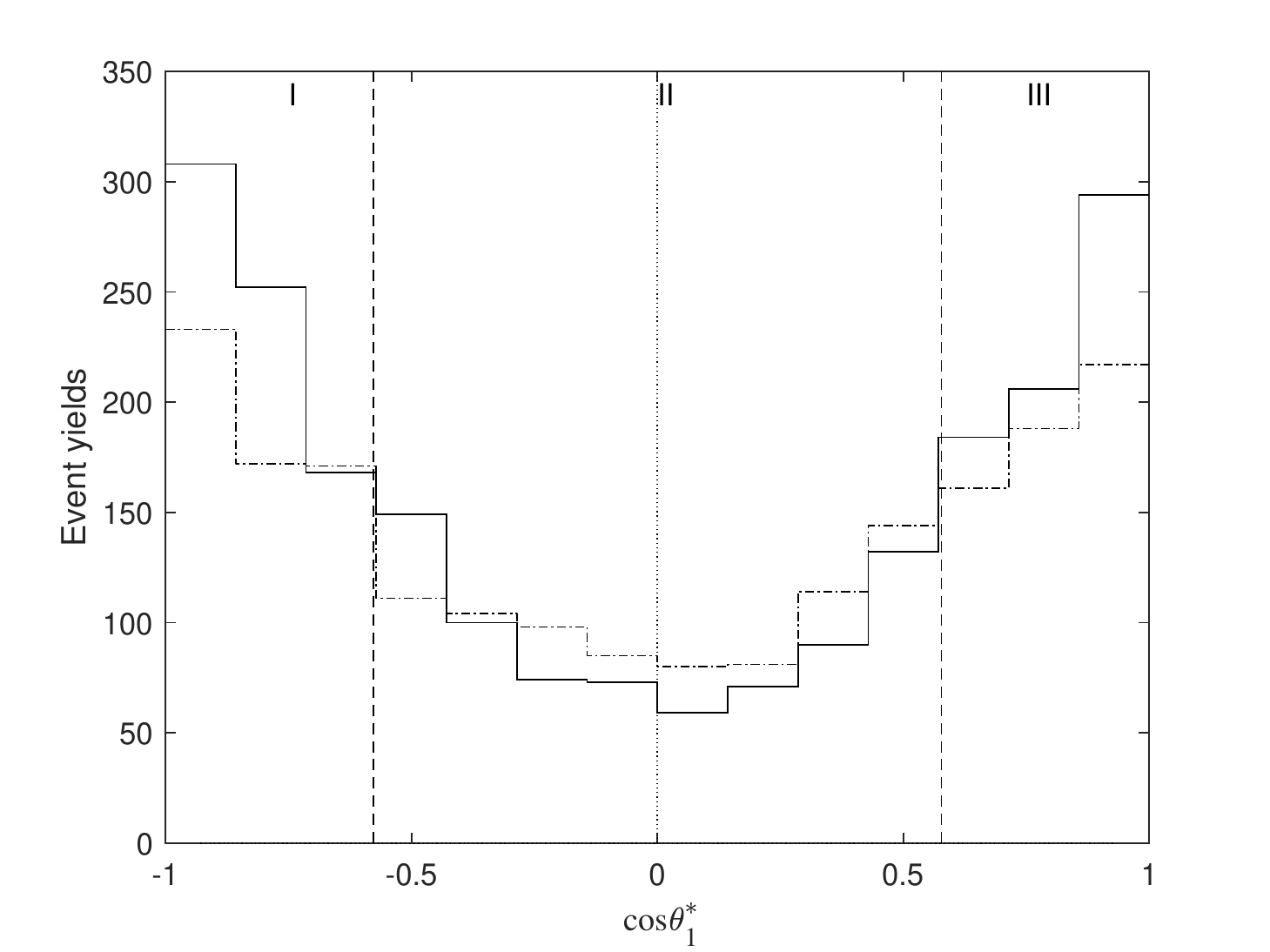}\\
  \caption{The simulation of the event yields for $\delta=4$. The solid line is for the baryon decay, while the dash-dotted line is for the anti-baryon decay.}\label{fig:EventSimu}
\end{figure}

In order to illustrate the advantage of the newly introduced PW$C\!P$As with respect to the significance and statistics, a simulation for the aforementioned decay process, $\Lambda_b^0\to p \pi^-\pi^+\pi^-$, for $\delta=4$, is performed.
The range of $c_{\theta^\ast_1}$ is uniformly divided into fourteen bins, so that the aforementioned three regions will approximately accommodate three (for Region I and III) or eight (for Region II) bins.
The distributions of the simulated event yields are presented in Fig. \ref{fig:EventSimu}.
In total, the event yields for the baryon and the anti-baryon decay are simulated to be 2202 and 1946, respectively.\footnote{
For comparison, the total signal yield for $\Lambda_b\to p \pi^-\pi^+\pi^-$ are $27600\pm200$ according to LHCb in Ref. \cite{Aaij:2019mmy}. When narrowed down to the phase space region on which we focus, our simulated event yields are reasonable. A reliable simulation of the event yield is unavailable because of various theoretical uncertainties.}
The values and uncertainties of various $C\!P$V observables calculated from the simulated data are list in Table \ref{tab:Simulate}, in which the uncertainties are assumed to be originated only from those of the event yields in each bin, which is estimated to be the square root of these event yields.
From this table, one can see clearly that the relative uncertainty of $A_{CP}^{(2)}$ is evidently reduced comparing with those of $A_{CP}^{(0)}$ and other r$C\!P$As.
In fact, according the simulation, the significance of $A_{CP}^{(2)}$  is $6.7\sigma$, comparing with the significance of other observables, which are $3.1\sigma$, $4.3\sigma$, $1.8\sigma$, and $3.2\sigma$, respectively.

\begin{figure}
  \centering
  \includegraphics[width=\textwidth]{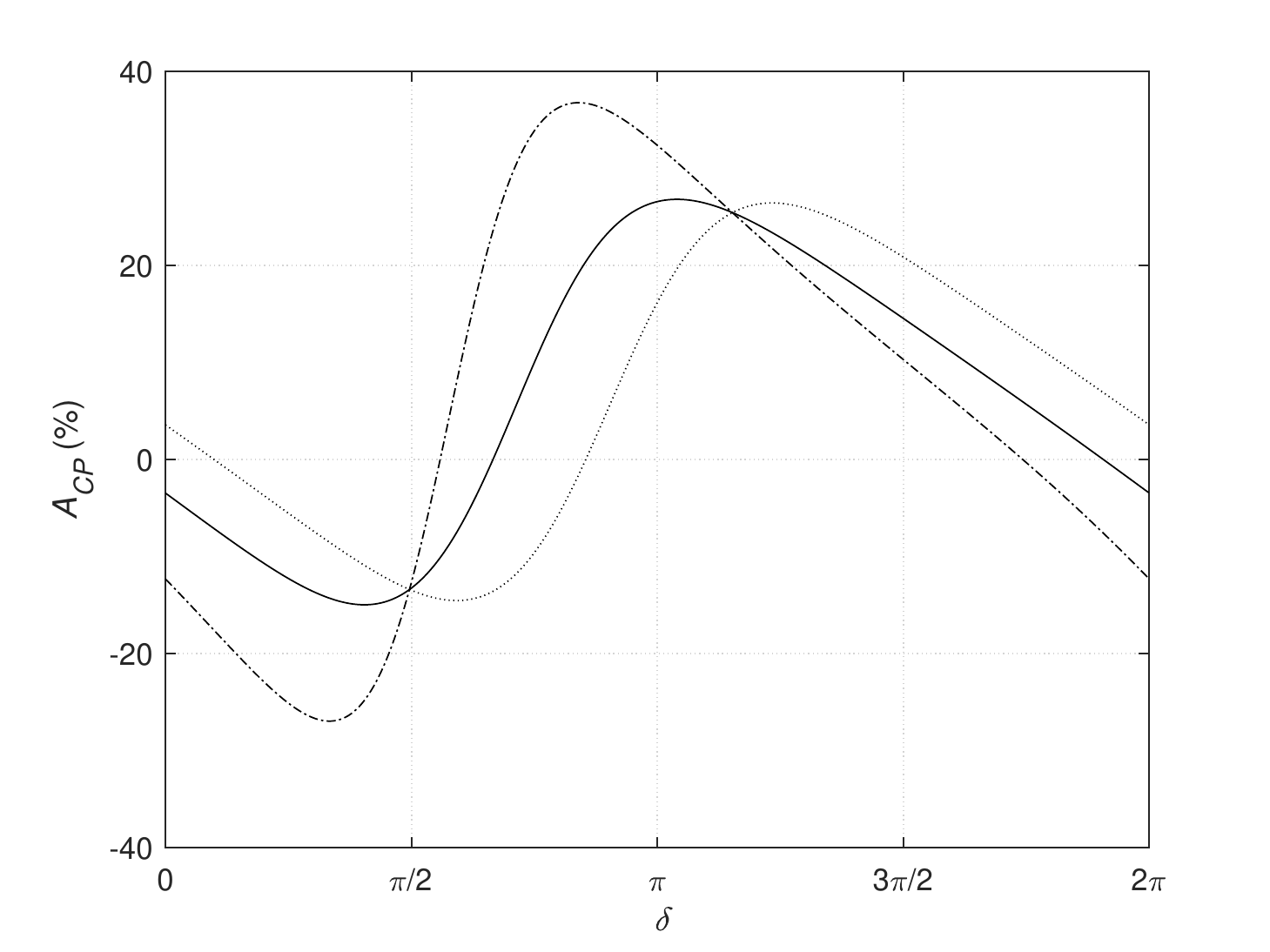}\\
  \caption{The PW$C\!P$As $A_{C\!P}^{(2)}$ defined in different intervals of $s_{p\pi^-}$ for $\Lambda_b^0\to p\pi^-\pi^+\pi^-$ near the resonance $\Delta^0(1232)$ as a function of the strong phase $\delta$. The solid line is for the situation that $s_{p\pi^-}$ is integrated in the region $((m_\Delta-\Gamma_\Delta)^2,(m_\Delta+\Gamma_\Delta)^2)$, the dotted line is for that $((m_\Delta-\Gamma_\Delta)^2,m_\Delta^2)$, and the dashed line is for that $(m_\Delta^2,(m_\Delta+\Gamma_\Delta)^2)$.}\label{fig:PWCPA3region}
\end{figure}

Another interesting observation from Fig. \ref{fig:PWCPA} is that for certain values of $\delta$, the $A_{C\!P}^{(2)}$ takes values very close to zero.
For example, when $\delta\approx 2.1$, $A_{C\!P}^{(2)}\approx0$.
This is mainly because the strong phase originated from the Breit-Wigner factor in the propagatror of $\Delta^0$, $1/(s_{p\pi^-}-m_\Delta^2+im_\Delta\Gamma_\Delta)$, varies rapidly when $s_{p\pi^-}$ is around the mass squared of $\Delta^0$.
As a consequence, when combined with certain values of $\delta$, large cancellation may come up, resulting in almost zero $A_{C\!P}^{(2)}$ when  $s_{p\pi^-}$ is integrated from $(m_\Delta-\Gamma_\Delta)^2$ to $(m_\Delta+\Gamma_\Delta)^2$.
To avoid this kind of cancellation, one just needs to consider the $A_{C\!P}^{(2)}$ defined in a different integration interval for $s_{12}$, for example, $s_{p\pi^-}\in((m_\Delta-\Gamma_\Delta)^2,m_\Delta^2)$, or $s_{p\pi^-}\in(m_\Delta^2,(m_\Delta+\Gamma_\Delta)^2)$.
Indeed, a simple calculation shows that $A_{C\!P}^{(2)}|_{s_{p\pi^-}\in((m_\Delta-\Gamma_\Delta)^2,m_\Delta^2)}\approx-13.5\%$ and $A_{C\!P}^{(2)}|_{s_{p\pi^-}\in(m_\Delta^2,(m_\Delta+\Gamma_\Delta)^2)}\approx23\%$ for $\delta=2.1$, as expected.
More details about the comparison among the $A_{C\!P}^{(2)}$'s defined in the aforementioned three intervals are illustrated in Fig. \ref{fig:PWCPA3region}.
On the experimental side, even if the search of $C\!P$V through the measurement of PWCPAs gained nothing for the interval from $(m_\Delta-\Gamma_\Delta)^2$ to $(m_\Delta+\Gamma_\Delta)^2$, one can change this interval for another try to avoid the potentially cancellation.

%%%%%%%%%%%%%%%%%%%%%%%%%
\section{\label{sec:conclusion}Conclusion}
%%%%%%%%%%%%%%%%%%%%%%%%%
$C\!P$V is now able to be investigated through multi-body decays of heavy hadrons such as $B$ meson, $D$ meson, bottom and charmed baryons.
In view of the yet negative outcome in searching for $C\!P$Vs in the baryon sector, we introduce a set of new observables, the PW$C\!P$As, which can be used to measure $CP$Vs in multi-body decays of heavy hadrons.
As explained in this work, the PW$C\!P$As provide a systematic and model-independent way in the investigation of $C\!P$As in multi-body decays of heavy hadrons.
We propose to search for $C\!P$Vs through the measurements of the PW$C\!P$As in multi-body decay channels of bottom and charmed baryons, such as $\Lambda_b^0\to p \pi^-\pi^+\pi^-$, $\Lambda_b\to p K^- \pi^+\pi^-$, $\Lambda^0_b\to p\pi^-K^+K^-$, $\Lambda_b^0\to\Lambda K^+\pi^-$, and $\Lambda_b^0\to p K_s \pi^-$.
It is possible that $C\!P$V in some multi-body decay channels of bottom and charmed baryons can be established with the strategy proposed in this paper.\footnote{
After the manuscript was accepted, we consider the case that the initial particle $H$ is polarized.
We find that our analysis in this work is also applicable to this case.
The main difference is that one need a non-trivial density matrix $\rho_{m_zm_z}^{H}$ ($m_z=-J,\cdots,J$) to describe the polarized initial state, so that the density matrix for the unpolarized initial state  $\sum_{m_z}Q_{H}^{(Jm_z)}$ is replaced by $\sum_{m_z}\rho^H_{m_zm_z}Q_{H}^{(Jm_z)}$ with $Q_{H}^{(Jm_z)}$ a dyad which is constructed by the initial state through $Q_{H}^{(Jm_z)}=\left|H(J,m_z)\rangle\langle H(J,m_z)\right|$.
Consequently, to repeat the analysis, one just need to add an extra factor $\rho_{m_zm_z}^{H}$ in Eq. (\ref{eq:wXY}), which satisfies 
$\rho_{m_zm_z}^{H}=\rho_{-m_z-m_z}^{\bar{H}}$ because of $C\!P$ symmetry in the $H$ and $\bar{H}$ production process.
The polarization of the initial state can also be described equivalently by a set parameters $\lambda_l^H$ ($l=0,1,\cdots,2J$), according to
$\sum_{m_z}\rho^H_{m_zm_z}Q_{H}^{(Jm_z)}=\sum_{l}\lambda_l Q_{H}^{(l)}$, with $Q_{H}^{(l)}=\sum_{m_z}\langle Jm_z J -m_z|l0\rangle Q_{H}^{(Jm_z)}$.
The relation between these two sets of parameters are easily obtained:
$\rho^H_{m_zm_z}=\sum_l \langle JJm_z ~ -m_z|l0\rangle \lambda^H_l$.
For spin-$1/2$ baryons, weather the initial state is polarized or not has nothing to do with PW$C\!P$A because of the rotational invariance.
}

\acknowledgments
 We thank Prof. Wen-Bin Qian for useful discussions. This work was supported by National Natural Science Foundation of China under Contracts Nos. 11705081 and 11775024.

\bibliography{zzh}

\begin{table}[h!]
  \begin{center}
    \caption{The values of $j$ for which the non-interference and interference terms appear in $w^{(j)}$ and $\bar{w}^{(j)}$ according to SRs in the cases of integer and half-integer spins of $R_a$ and $R_b$.}
    \label{tab:IntHalfIntSR}
\begin{tabular}{c|c|c||c|c|c}
  \hline
   $\left({J^P}_{R_a},{J^P}_{R_b}\right)$ & non-int. $j$ & int. $j$ &$\left({J^P}_{R_a},{J^P}_{R_b}\right)$ & non-int. $j$ & int. $j$  \\
  \hline
  $\left(0^{+},0^{+}\right)$ or $\left(0^{-},0^{-}\right)$ & 0 & 0 & $\left(\frac{1}{2}^{+},\frac{1}{2}^{+}\right)$ or $\left(\frac{1}{2}^{-},\frac{1}{2}^{-}\right)$ & 0 & 0\\
  \hline
  $\left(0^{+},0^{-}\right)$ or $\left(0^{-},0^{+}\right)$ & 0 & none  & $\left(\frac{1}{2}^{+},\frac{1}{2}^{-}\right)$ or $\left(\frac{1}{2}^{-},\frac{1}{2}^{+}\right)$ & 0 & 1 \\
  \hline
  $\left(0^{+},1^{+}\right)$ or $\left(0^{-},1^{-}\right)$  &  0, 2 &  none &
  $\left(\frac{1}{2}^{+},\frac{3}{2}^{+}\right)$ or $\left(\frac{1}{2}^{-},\frac{3}{2}^{-}\right)$  &  0, 2 &  2  \\
  \hline
  $\left(0^{-},1^{+}\right)$ or $\left(0^{+},1^{-}\right)$ &  0, 2 & 1 & $\left(\frac{1}{2}^{-},\frac{3}{2}^{+}\right)$ or $\left(\frac{1}{2}^{+},\frac{3}{2}^{-}\right)$ &  0, 2 & 1 \\
  \hline
  $\left(0^{+},2^{+}\right)$ or $\left(0^{-},2^{-}\right)$  & 0, 2, 4 &  2 & $\left(\frac{1}{2}^{+},\frac{5}{2}^{+}\right)$ or $\left(\frac{1}{2}^{-},\frac{5}{2}^{-}\right)$ &  0, 2, 4 & 2 \\
  \hline
  $\left(0^{-},2^{+}\right)$ or $\left(0^{+},2^{-}\right)$ &  0, 2, 4 & none &   $\left(\frac{1}{2}^{-},\frac{5}{2}^{+}\right)$ or $\left(\frac{1}{2}^{+},\frac{5}{2}^{-}\right)$ &  0, 2, 4 & 3\\
  \hline
  $\left(1^{+},1^{+}\right)$ or $\left(1^{-},1^{-}\right)$ &  0, 2 & 0, 2 &  $\left(\frac{3}{2}^{+},\frac{3}{2}^{+}\right)$ or $\left(\frac{3}{2}^{-},\frac{3}{2}^{-}\right)$ &  0, 2 & 0, 2 \\
  \hline
  $\left(1^{-},1^{+}\right)$ or $\left(1^{+},1^{-}\right)$ &  0, 2 & 1 & $\left(\frac{3}{2}^{-},\frac{3}{2}^{+}\right)$ or $\left(\frac{3}{2}^{+},\frac{3}{2}^{-}\right)$ &  0, 2 & 1, 3  \\
  \hline
  $\left(1^{+},2^{+}\right)$ or $\left(1^{-},2^{-}\right)$ &  0, 2, 4 & 2 &   $\left(\frac{3}{2}^{+},\frac{5}{2}^{+}\right)$ or $\left(\frac{3}{2}^{-},\frac{5}{2}^{-}\right)$ &  0, 2, 4 & 2, 4 \\
  \hline
  $\left(1^{-},2^{+}\right)$ or $\left(1^{+},2^{-}\right)$ &  0, 2, 4 & 1, 3  & $\left(\frac{3}{2}^{-},\frac{5}{2}^{+}\right)$ or $\left(\frac{3}{2}^{+},\frac{5}{2}^{-}\right)$ &  0, 2, 4 & 1, 3 \\
  \hline
  $\left(2^{+},2^{+}\right)$ or $\left(2^{-},2^{-}\right)$ &  0, 2, 4 & 0, 2, 4 &   $\left(\frac{5}{2}^{+},\frac{5}{2}^{+}\right)$ or $\left(\frac{5}{2}^{-},\frac{5}{2}^{-}\right)$ &  0, 2, 4 & 0, 2, 4 \\
  \hline
  $\left(2^{-},2^{+}\right)$ or $\left(2^{+},2^{-}\right)$ &  0, 2, 4 & 1, 3  & $\left(\frac{5}{2}^{-},\frac{5}{2}^{+}\right)$ or $\left(\frac{5}{2}^{+},\frac{5}{2}^{-}\right)$ &  0, 2, 4 & 1, 3, 5 \\
  \hline
\end{tabular}
  \end{center}
\end{table}

\begin{table}[h!]
  \begin{center}
    \caption{The simulated values and uncertainties of various observables (in \%), for $\delta=4$. The theoretical values are also presented for comparison.}
    \label{tab:Simulate}
\begin{tabular}{c|c|c|c|c|c}
  \hline
   $\delta=4$&$A_{CP}^{(2)}$ & $A_{CP}^{(0)}$ & $A_{CP}^{{\rm Reg~I}}$ & $A_{CP}^{{\rm Reg~II}}$ & $A_{CP}^{{\rm Reg~III}}$ \\
  \hline
  Simulation&$25.6\pm3.8$ & $4.9\pm1.6$ & $11.7\pm2.7$ & $-4.4\pm2.5$ & $9.0\pm2.8$ \\
  \hline
  Theory&$21.8$ & $6.8$ & 11.9  & $-1.5$   &  11.7
  \\  \hline
\end{tabular}
  \end{center}
\end{table}


\end{document}